\documentstyle[11pt]{article}
\renewcommand{\baselinestretch}{1.3}

\input epsf

\begin{document}

\begin{titlepage}
\renewcommand{\thefootnote}{\fnsymbol{footnote}}
\renewcommand{\baselinestretch}{1.3}
\hfill UWThPh - 1996 - 3
\medskip
\vfill
\begin{center}
{\LARGE {String Supported Wormhole Spacetimes\\
and Causality Violations}}
\medskip
\vfill
\renewcommand{\baselinestretch}{1}
{\large F. SCHEIN 
\footnote{e-mail: schein@pauli.thp.univie.ac.at\newline
\hspace*{8 pt}
$^{**}$e-mail: pcaich@pap.univie.ac.at \newline
\hspace*{8 pt}
$^\diamondsuit$e-mail: israel@euclid.phys.ualberta.ca }
and  P.C. AICHELBURG$^{**}$\\ 
\medskip
Institut f\"ur Theoretische Physik \\
Universit\"at Wien \\
Boltzmanngasse 5, A--1090 Wien, \\
Austria \\}
\medskip \medskip
{\large {W. ISRAEL$^\diamondsuit$}\\
\medskip
Canadian Institute for Advanced Research Cosmology Program, \\
Theoretical Physics Institute, University of Alberta, \\
Edmonton, Canada T6G 2J1} 
\end{center}
\vfill

\begin{abstract}
We construct a static axisymmetric wormhole from the gravitational field of 
two Schwarzschild particles which are kept in equilibrium by  strings 
(ropes) extending to infinity. The wormhole is obtained by matching two
three-dimensional timelike surfaces surrounding each of the particles and 
thus spacetime becomes non-simply connected. Although the matching will
not be exact in general it is possible to make the error arbitrarily small by
assuming that the distance between the particles is much larger than the 
radius of the wormhole mouths. Whenever the masses of the two wormhole
mouths are different, causality violating effects will occur.
\end{abstract}

\end{titlepage}
 
\section{Introduction}
The field equations of Einsteins theory, being local, do not fix the global
structure of spacetime. Kurt G\"odel \cite{goedel} noted first in 1949 that
general relativity admits topologically nontrivial solutions exhibiting causal
paradoxes, including closed timelike curves. Whatever one's attitude to these
paradoxes - whether one views them simply as embarrassments, as a spur to
re-examining the foundation of physics, as a challenge to philosophers of
physics, or as a clue to new physical principles which might outlaw or tame the
pathologies - the issues they raise are non-ignorable and have recently
received a good deal of attention [2-5].
 	
In 1988, M. Morris and K. Thorne \cite{MorrisThorne} conceived the idea that
an advanced civilization might be able to construct a traversible wormhole
which connects
two distant regions of space. To prevent the throat from closing, exotic
material (with negative energy-density) needs to be packed into the hole. Such
a wormhole can be employed as a time-machine by setting the two mouths into
high speed relative motion.
V.Frolov and I.Novikov \cite{NovikovFrolov} pointed out in 1990 that even a
static wormhole
functions as a time-machine if the two mouths are at different gravitational
potentials, e.g. if one of the mouths is held near a neutron star. These are
probably the simplest models for illustrating and testing ideas about
causality-violating spacetimes.
    	
The two mouths appear as two separate masses in the asymptotically flat exterior
space. Spacetime can therefore not be spherically symmetric. Moreover,
the mouths will naturally gravitate towards each other, thus violating the
assumed staticity. The nonstaticity can be made arbitarely small for sufficient
large initial separation of the mouths. Alternatively, it can be
eliminated altogether by anchoring the mouths, e.g. by strings held at
infinity. It is the aim of this paper to develop an explicit, strictly static,
wormhole model of this type.
    	
In Section II we start from the so-called Bach-Weyl solution which describes
two
static particles held in equilibrium by a strut. The strut may be replaced by
strings that are attached to each of the particles and run off to infinity. In
Sec.III we construct a wormhole from the Bach-Weyl solution by cutting out the
interior of two timelike tubes each surrounding one of the particles. Matching 
these boundaries gives rise to a layer of (non-standard) matter. The
form of the boundaries is described by two "deformation-functions" which are
required to make the induced metric continuous. We show that solutions for the
deformation-functions exist (Appendix A) and give their explicit form for the
case that the two wormhole mouths are far apart. 
However, this yields only two out of
the three components of the induced metric continuous. To smooth out the
remaining metric component we introduce additional matter. The result is that
this does not significally contribute to the stress-energy tensor as long as
the mouths are sufficiently separated. In Sec.IV we calculate the stress-energy
tensor for the surface layer. Finally, we point out that such wormholes may be
used as a time-machine whenever the mass parameters of the two wormhole mouths
are different. 
    	
\section{The Bach-Weyl-Solution}
In this section we introduce the so-called Bach-Weyl solution which describes 
two static Schwarzschild particles held in equilibrium by a strut between 
them.  For this we follow the work of W. Israel and K.A. Khan \cite{IsraelKhan}
who calculated the gravitational field of arbitrarily many collinear particles.
Any static axially symmetric spacetime can be written in
a cylindrical coordinate system $\tilde I (T,r,z,\varphi)$
in the form
\begin{equation} \label{metric}
ds^{2}=-e^{2\lambda(r,z)}dT^{2}+e^{2(\nu(r,z)-\lambda(r,z))}(dr^{2}+dz^{2})
       +r^{2}e^{-2\lambda(r,z)}d\varphi^{2}
\end{equation}
depending on the two functions $\nu(r,z)$ and $\lambda(r,z)$. 
The vacuum Einstein equations require $\lambda(r,z)$ to be an 
axisymmetric Newtonian
potential function of $(r,z,\varphi)$, treated as cylindrical coordinates on a
fictitious flat background.
It is known that if the Schwarzschild line element of a particle with mass $m$
is transformed
to the coordinate system $\tilde I$ , $\lambda$ is formally the Newtonian
potential of a uniform rod with length $2m$. 
Therefore, we may use the following picture:
Consider two non-overlapping rods placed along the z-axis, their centers
located at $z=a_{i}$ (i=1,2). We assume that the $i^{th}$ rod has
mass $b_{i}$ and define the quantities $\rho_{i}^{\pm}(r,z)$
which measure the distance from its ends (see Fig.1):
\begin{eqnarray} \label{def}
(\rho_{i}^+)^{2}=r^{2}+(z_{i}^+)^{2}, \qquad 
 (\rho_{i}^-)^{2}=r^{2}+(z_{i}^-)^{2} \nonumber\\
z_{i}^+=z-(a_{i}+b_{i}), \qquad
z_{i}^-=z-(a_{i}-b_{i})    
\end{eqnarray}
Following \cite{IsraelKhan} we introduce the notation
\begin{eqnarray}
E(i^+,j^+)=\rho_{i}^+\rho_{j}^++z_{i}^+z_{j}^++r^{2} \nonumber\\
E(i^-,j^+)=E(j^+,i^-)=\rho_{i}^-\rho_{j}^++z_{i}^- z_{j}^++r^{2}
\end{eqnarray}
with an analogous expression defining $E(i^-,j^-)$.
As a solution of the Einstein equations for our two body problem 
we choose $\lambda$ to be the Newtonian potential of these rods:
\begin{equation} \label{lam}
\lambda(r,z)= \lambda_{1}(r,z)+\lambda_{2}(r,z), \qquad 
\lambda_{i}=\frac{1}{2}
            \ln\frac{\rho_{i}^++\rho_{i}^--2b_{i}}{\rho_{i}^++\rho_{i}^-+2b_{i}}
\end{equation}
The corresponding $\nu(r,z)$ is given by the sum
\begin{equation} \label{nu}
\nu(r,z)=\sum_{i=1}^{2}\sum_{j=1}^{2}\nu_{ij}(r,z),\qquad \mbox{where} \qquad
\nu_{ij}=\frac{1}{4}\ln\frac{E(i^-,j^+)E(i^+,j^-)}{E(i^+,j^+)E(i^-,j^-)}.
\end{equation}     
With this choice the constant of integration has been adjusted to make $\nu$ 
vanish at spatial infinity. Thus the spacetime is asymptotically flat and 
satisfies the vacuum equations everywhere except on the segment of the 
$z$-axis between the the two rods.
   	\begin{equation}
\nu_{0}=\nu(r=0,z)=\frac12\ln\vert\frac{(a_{1}-a_{2})^2-(b_{1}+b_{2})^2}
   {(a_{1}-a_{2})^2-(b_{1}-b_{2})^2}\vert, \qquad z \in [a_1+b_1,a_2-b_2]
\end{equation}
This phenomenon was interpreted by Bach and Weyl as a strut which holds the
two bodies apart. In \cite{Israel} one of the authors has considered
how a string energy 
density and tension can be defined for such line singularities.
For our case the non-vanishing 
components of the energy-momentum tensor $L_{\mu}^{\nu}$ of the line source are
given by:
\begin{equation}
L_{0}^{0}=L_{2}^{2}=\frac{1}{4}(1-e^{\nu_{0}}) \qquad \geq 0 \qquad if
                                               \qquad b_{1},b_{2}\geq 0
\end{equation}
The strut therefore has a negative energy density and a pressure numerically 
equal to it.
Another possibility for keeping the two particles apart which 
avoids a strut with
negative energy density is offered by a different choice for the constant of 
intergration in equation (\ref{nu}). If e.g. we choose $\nu=-\nu_{0}$, our 
solution would be regular on the segment of the $z$-axis
between the two rods. On the other hand,  there are now two singularities  
extending from the two rods to spatial infinity. Staticity requires that 
these "strings" have tension and positive energy density.
   
\section{Matching of  Wormhole  Mouths}
   	
We cut out the interior of the surfaces $S_{1}$ and $S_{2}$ 
surrounding each of the two particles - the wormhole mouths -
and match  the surfaces to get
a non-simply connected spacetime. Let us introduce two spherical polar 
coordinate systems $I$ and $II$, one centered at \(z=a_1\) the other at
\(z=a_2\) (see Fig.1):
\begin{eqnarray}\label{tran} 
I(t,r_{1},\vartheta,\varphi):\qquad \qquad \qquad t=e^{\Lambda_{1}}T \nonumber\\
                 r=\sqrt{r_{1}(r_{1}-2b_{1})}\sin{\vartheta}\nonumber\\
                        z=a_{1}+(r_{1}-b_{1})\cos{\vartheta} \\
II(t,r_{2},\vartheta,\varphi):\qquad \qquad \qquad t=e^{\Lambda_{2}}T 
        \nonumber\\
   	              r=\sqrt{r_{2}(r_{2}-2b_{2})}\sin{\vartheta} \nonumber\\ 
   	              z=a_{2}+(r_{2}-b_{2})\cos{(\pi-\vartheta)} 
\end{eqnarray}
We do not distinguish time and angle coordinates of the different coordinate
patches because  we want to identify points on the wormhole mouths 
with equal values of $(t,\vartheta,\varphi)$. The constants $\Lambda_{i}$
determine the transformation of the time
coordinate of the metric (\ref{metric}) to the charts I and II. We
will see that they cannot be equal in general which is the
crucial fact leading to the occurence of closed
timelike curves (CTCs).
Notice that the polar axis 
of the spherical coordiate system near mouth $S_1$ points to
the $+z$ - direction but the polar axis of the  spherical 
coordinate system near 
mouth $S_2$ points to the $-z$ - direction. With this
choice we identify the inner poles $A$ and $B$ of the wormhole mouths (see
Fig.1).
   	
In these coordinates the Bach-Weyl-solution becomes
\begin{eqnarray} \label{bach}
(ds^2)_{j}=e^{2(\sigma_{j}-\lambda_{i})}\lbrace
\frac{dr^2_{j}}{(1-\frac{2b_{j}}{r_{j}})}+r^2_{j}d\vartheta^2\rbrace+
e^{-2\lambda_{i}}r^2_{j}\sin{\vartheta}^2d\varphi^2 \nonumber\\
-e^{2(\lambda_{i}-\Lambda_{j})}(1-\frac{2b_{j}}{r_{j}})dt^2,
                                          \qquad i,j=1,2 \quad ; \quad i\not=j
\end{eqnarray}
where $\sigma_{i}$ is defined as the perturbation of the function $\nu(r,z)$
induced by the other particle, i.e.
\begin{equation}\label{sig}
\nu=\nu_{11}+\nu_{22}+2\nu_{12}=\nu_{11}+\sigma_{1}=\nu_{22}+\sigma_{2}
\end{equation}
Now let the parametric equations of the surfaces $S_1$ and $S_2$ in our charts
$I$ and $II$ be $x_{i}^{\mu}=x_{i}^{\mu}(\xi^a)$ with three-dimensional
intrinsic coordinates $\xi^a=(t,\vartheta, \varphi)$ - notice that $i,j=1,2
\quad \mbox{but} \quad a,b=0,2,3$:  
\begin{eqnarray}
S_{1}: \qquad r_{1}=Re^{\delta(\vartheta)} \nonumber\\ 
S_{2}: \qquad r_{2}=Re^{\epsilon(\vartheta)}
\end{eqnarray}
The deformation functions $\epsilon(\vartheta)$ and $\delta(\vartheta)$ 
just depend on the
angle $\vartheta$ and will be determined in the following.
First we calculate the induced metric on $S_1$ and $S_2$.
The three
holonomic basis vectors $e_{(a)}\vert_{S_i}=\frac{\partial}{\partial
\xi^a}\vert_{S_i}$
tangent to $S_i$ have components
$e_{(a)}^\mu \vert_{S_i}=\frac{\partial x_{i}^
{\mu}}{\partial \xi^a}$ and their scalar products define the metric induced 
on $S_i$:
\begin{equation}
g_{ab}\vert_{S_i}=e_{(a)}\vert_{S_i} \cdot e_{(b)} \vert_{S_i}
\equiv g_{\alpha \beta} e_{(a)}^{\alpha} e_{(b)}^{\beta} \vert_{S_i}
\end{equation}
which for $S_1$ reads:
\begin{eqnarray} \label{ind}
(ds^2)_{S_{1}}=e^{2(\sigma_{1}-\lambda_{2}+\delta)}R^2
\lbrace \frac{(\frac{d\delta}{d\vartheta})^2}
{(1-\frac{2b_1}{Re^{\delta}})}+1 \rbrace d\vartheta^2 + \nonumber\\
e^{-2\lambda_{2}}R^2e^{2\delta(\vartheta]}\sin^2{(\vartheta)} d\varphi^2 
-e^{2(\lambda_{2}-\Lambda_{1})}(1-\frac{2b_{1}}{Re^{\delta}})dt^2
\end{eqnarray}
To get the induced metric on $S_2$ one has to interchange the indexes 1 and 2 
and to replace $\delta$ by $\epsilon$.
Next, we match these surfaces by identifying points with equal values of
$(t,\vartheta,\varphi)$, i.e. $S_1 \equiv S_2 \equiv S$.
To obtain a continuous four-metric the induced metric has to be
the same on $S_1$ and $S_2$.
We may achieve continuity
on $S$ for the metric coefficients $g_{00}$ and $g_{33}$  by specifying 
the functions $\delta(\vartheta)$ and $\epsilon(\vartheta)$ and the relation
between the asymptotic time scales,
\begin{equation}
C:=e^{\Lambda_1-\Lambda_2}.
\end{equation}
In general, continuity for $g_{22}$ can not be achieved exactly. However,
we will see that the error in the $g_{22}$ component can be made
arbitrarily small by enlarging the distance $D=a_2-a_1$ between the
wormhole mouths.

Throughout this paper we assume that the mass parameters $b_i$ are positive.
Hence, to make sure that the surfaces $S_i$ are timelike we impose the
condition  
\begin{equation} \label{appr}
2b_{1,2} < R \ll D
\end{equation}
In Appendix A we show that under this condition also the radii 
$Re^{\delta}$ and $Re^{\epsilon}$
are greater than the Schwarzschild-radii $2b_1$
and $2b_2$ respectively.

To calculate the deformation functions $\delta$ and $\epsilon$ we 
equate the metric-coefficients
$g_{33} \vert_{S_{1,2}}$ and $g_{00} \vert_{S_{1,2}}$ which 
yields a system of two equations:
\begin{equation} \label{deleps}
-\lambda _1(\epsilon,\vartheta)+\epsilon(\vartheta)=
-\lambda _2(\delta,\vartheta)+\delta(\vartheta) 
\end{equation}
\begin{equation} \label{g00}
e^{2(\lambda_{2}-\lambda_{1})}\left(1-\frac{2b_{1}}{Re^{\delta}}\right)=
C^2\left(1-\frac{2b_{2}}
{Re^{\epsilon}}\right)
\end{equation} 
There are different possibilties for
fixing the size of the wormhole mouths, e.g. by choosing $\delta$ at an
arbitrary point on the surface $S_1$ or by choosing the constant $C$.
Notice that if the masses $b_1$ and $b_2$  are
equal, exact matching of the two wormhole mouths is possible. In this case 
it follows from equations (\ref{deleps}) and (\ref{g00})  that
$\delta(\vartheta)=\epsilon(\vartheta)\nonumber$ and $C=1$.
Such a spacetime will be symmetric with respect to the plane 
$z=\frac12(a_2-a_1)$ 
and will not contain a Cauchy horizon. The identification
of the mouths can be chosen in such a way that either CTCs exist throughout
the whole spacetime - an eternal time machine - or that there do not exist 
CTCs at all.
Under the condition (\ref{appr}) we can show that if 
$b_1 \not= b_2$ then $C \not= 1$ in order that a solution to the system
of equations
(\ref{deleps}) and (\ref{g00}) exists. Hence, although our constructed
spacetime is
static, there does not exist a global timelike Killing vector field.
In the terminology of V.Frolov and I.Novikov \cite{NovikovFrolov} the
gravitational field is nonpotential and according to their general proof
CTCs have to occur.  V.Frolov and I.Novikov studied explicitly 
a spherically symmetric wormhole model with two asymptotic regions. 
In order to be able to compare our results with theirs we choose $C$ 
to have the value 
\begin{equation} \label{cee}
C=\sqrt{\frac{\left( 1-\frac{2b_1}{R}\right) }
{\left( 1-\frac{2b_2}{R}\right)}}.
\end{equation}
With this choice of the constant $C$ and under the assumptions (16)
we are able to show that a solution  $\delta$ and $\epsilon$ to the system 
(\ref{deleps},\ref{g00}) exists for every $\vartheta$.
To calculate the deformations we eliminate $(\lambda_2-\lambda_1)$ from eqn.
(\ref{g00}) by using eqn. (\ref{deleps}) and solve for $\delta$:
\begin{equation}\label{delta}
e^{\delta}=\frac{b_1}{R}+\sqrt{\left(\frac{b_1}{R}\right)^2+
C^2\left(e^{2\epsilon}-\frac{2b_2}{R}e^\epsilon\right)}
\end{equation}
Putting this result back into eqn. (\ref{deleps}) leads to an 
implicit equation for $\epsilon(\vartheta)$ which cannot be solved exactly:
\begin{equation}\label{ns} 
F_D(\vartheta,e^{\epsilon}):=e^{\epsilon}-e^{\lambda_{1}(\epsilon)-
\lambda_{2}(\delta)}e^{\delta(\epsilon)}=0
\end{equation} 
However, if he wormhole mouths are infinitly apart i.e. $D$ to infinity
eq.(\ref{ns}) reduces to
\begin{equation}\label{nsinf} 
F_{\infty}(\vartheta,e^{\epsilon}):=e^{\epsilon}-e^{\delta(\epsilon)}=0
\end{equation} 
By virtue of equations (\ref{cee}) and 
(\ref{delta}) one checks that $\epsilon$ = 0 is a solution. 
This is to be expected since deviations from 
spherical symmetry should tend to zero as the 
distance increases.
In Appendix A we show that for sufficiently large $D$ a 
solution to eq.(\ref{ns}) 
always exists. Moreover it is shown that the solution is well approximated
by Newton`s method starting with $\epsilon=0$ as zeroth approximation. 
To first order in $R$/$D$ the solution is:
\begin{equation}\label{eps}
e^{\epsilon(\vartheta)} \simeq 1-\frac{R}{D}(1-\frac{2b_2}{R})(1-\frac{b_1}{R})
\end{equation}
An analogous result holds for the function $e^{\delta(\vartheta)}$ with
the mass-parameters $b_1$ and $b_2$ interchanged. 
We see that the wormhole mouths are slightly deformed spheres
of radius $R$ and therefore our spacetime may be directly compared to
the wormhole spacetimes treated in \cite{NovikovFrolov}. 
   
Now we focus attention on the metric component $g_{22}$. We have to
check the behaviour of the quantities $\frac{d\delta}{d\vartheta}$,
$\frac{d\epsilon}{d\vartheta}$ and $\sigma_i$ under our assumption
(\ref{appr}). Using the implicit function theorem and expanding in powers of 
$\frac{R}{D}$ (see Appendix A) we find
\begin{equation}\label{deriv}
\frac{d\delta}{d\vartheta} , \frac{d\epsilon}{d\vartheta}
\sim O \left(\frac{R^2}{D^2} \right) 
\end{equation}

The discontinuity of the induced metric component $g_{22}$ (see eqn.(\ref{ind}))
depends on the angle $\vartheta$ and is given by:
\begin{eqnarray}
g_{22}\vert_{S_2}-g_{22}\vert_{S_1}=
e^{2(\sigma_1(\vartheta)-\lambda_2(\vartheta))}R^2e^{2\delta}
\triangle g_{22}(\vartheta) \nonumber\\
\mbox{where} \quad \triangle g_{22}(\vartheta):=
\left[ e^{2(\sigma_{2}(\vartheta)-\sigma_1({\vartheta}))}
\lbrace \frac{(\frac{d\epsilon}{d\vartheta})^2}
{(1-\frac{2b_2}{Re^{\epsilon}})}+1 \rbrace - 
\lbrace \frac{(\frac{d\delta}{d\vartheta})^2}
{(1-\frac{2b_1}{Re^{\delta}})}+1 \rbrace \right] 
\end{eqnarray}
Using the definition of the functions $\sigma_i(r,z)$ given by (\ref{sig}), 
inserting the functions $v_{ij}$ from equation (\ref{nu}) and expanding this in
powers of $R/D$ yields the approximation:
\begin{equation}
\triangle g_{22}(\vartheta) \sim \frac{-4(b_2-b_1)b_1b_2}{D^3} sin^2\vartheta 
+O\left(\frac{R^4}{D^4}\right)
\end{equation}
Thus, for given $b_1/R$ and $b_2/R$ the discontinuity 
can be made arbitrarily small by choosing the distance $D$ to be
suitably large without changing the constant $C$.
Causality violating effects will be unaffected by the distance
of the wormhole mouths.
 
The price for constructing a transversable wormhole is of course that 
non-classical matter must be present. In our case there is an infinitely thin 
shell concentrated at the surface $S$. 
Because of the discontinuity of the metric on this surface 
the Einstein-tensor is not well defined there, and in principle we cannot apply
the thin shell formalism to calculate the stress-energy tensor. By 
a small modification of the metric - which physically means that we introduce
some additional matter on
one side of the shell - we circumvent this problem without strongly influencing 
the results obtained by following the standard shell formalism.
We add a perturbation term to the component $g_{\vartheta \vartheta}$
of the Bach-Weyl metric (\ref{bach}) in chart $I$ which should exactly cancel 
the discontinuity of the induced metric on the surface $S$:
\begin{equation}\label{permetric}
g_{\vartheta \vartheta}(r_1,\vartheta) \to
e^{2(\sigma_1(r_1,\vartheta)-\lambda_2(r_1,\vartheta))}r_1^2 
\lbrace 1+\triangle g_{22}(\vartheta)
f_{\alpha}(r_1-Re^{\delta(\vartheta)})\rbrace
\end{equation}
where we have introduced a profile function $f_{\alpha}(x)$.
This function should be at 
least twice differentiable and satisfy the properties 
$f_\alpha(0)=1$ and $f_\alpha(x)=0$ for $x \geq \alpha$.
For example we may take it to be of the form:
\begin{eqnarray}\label{sugg}
f_\alpha(x)=e^{\frac{-\alpha^2}{(\alpha^2-x^2)}+1} 
                    \quad \mbox{for} \quad \vert x \vert < \alpha \nonumber\\
                    = 0 \quad \mbox{for} \quad  \vert x \vert \geq \alpha
\end{eqnarray}
The stresses and the energy associated with the additional
matter surrounding the shell can be analysed by
calculating the Einstein-tensor of the new perturbed metric. 
As expected, it turns out that the energy-
momentum tensor is of the same order as the perturbation 
function, i.e $O(R^3/D^3)$, 
but also depends on the first and second derivatives of the
profile function which are proportional to $1/\alpha$ 
and $1/\alpha^2$ respectively. 
Therefore the matter distribution has to be spread over 
a region thicker than $R/D$ to keep its
density low. Notice that in general this matter will also 
violate the energy conditions.

\section{Stress - Energy of the Wormhole Shell}
The surface stress-energy tensor $S_{ab}$ of the
layer is linked to the jump  $[K_{ab}]:=K_{ab}^{(II)}-K_{ab}^{(I)}$ 
of normal extrinsic curvature across $S$ \cite{BarrabesIs}.
$K_{ab}^{I,II}$ are the extrinsic curvatures corresponding to the different
imbeddings of $S$ in charts I and II, each defined as
\begin{equation}
K_{ab}:=-n \cdot \nabla_{e_{(b)}} e_{(a)}
\end{equation}
The vector $n$ is the normal to the surface $S$ pointing
from chart $I$ to $II$ and is normalized to one, $e_{(a)}$ are the three
holonomic basis vectors of the surface $S$ defined above. 
For non-lightlike surfaces the following distributional equivalent of 
Einstein's field equations holds: 
\begin{equation} \label{ein}
-8\pi (S_{ab}-\frac 12g_{ab}S)= \left[ K_{ab} \right] 
                              = K_{ab}^{(II)}-K_{ab}^{(I)}
\end{equation}
Notice that the stress-energy-tensor does not depend on the sign of 
the normal vector $n$ because if it would point to the opposite direction, the
two terms on the right hand side of equation (\ref{ein}) would have to be
interchanged.
For the unperturbed metric (\ref{bach}) the components of $n$ with respect to 
the different charts are:
\begin{equation}
n_{\mu}^{(I)}=\left(0,-1,
\frac{1}{Re^{\delta}}\frac{d\delta}{d\vartheta},0\right)
\frac{e^{(\sigma_1-\lambda_2)}}{\sqrt{1-\frac{2b_1}{Re^\delta}+
\left(\frac{d\delta}{d\vartheta}\right)^2}}
\end{equation}
\begin{equation}
n_{\mu}^{(II)}=\left(0,1,
-\frac{1}{Re^{\epsilon}}\frac{d\epsilon}{d\vartheta},0\right)
\frac{e^{(\sigma_2-\lambda_1)}}{\sqrt{1-\frac{2b_2}{Re^\epsilon}+
\left(\frac{d\epsilon}{d\vartheta}\right)^2}}
\end{equation}
As expected the calculation yields 
that the energy density of the wormhole mouth measured by any 
observer will be negative, pressures will be positive:
\begin{eqnarray}
S_0^0 \simeq \frac{1}{4\pi R}
\left(\sqrt{1-\frac{2b_1}{R}}
+\sqrt{1-\frac{2b_2}{R}}\right)+O(\frac{R}{D})
\nonumber\\
S_2^2=S_3^3 \simeq \frac{1}{8\pi R}
\left(\frac{1-\frac{b_1}{R}}{\sqrt{1-\frac{2b_1}{R}}}
+\frac{1-\frac{b_2}{R}}{\sqrt{1-\frac{2b_2}
{R}}}\right)+O(\frac{R}{D})
\end{eqnarray} 
If we apply the thin shell formalism to the perturbed spacetime
(\ref{permetric}) the result for the stress-energy tensor 
of the shell does not differ from the calculation 
using the original metric up to order $O(R^3/D^3)$. 
This results from the fact that the perturbation in the 
metric and its first derivative 
with respect to $\vartheta$ are of this order. 
(For the profile function (\ref{sugg}) 
the first derivative with respect to the $r_1$ vanishes on $S$.)  

\section{Conclusion}
V. Frolov and I. Novikov \cite{NovikovFrolov} considered the general 
situation of static wormhole spacetimes where the gravitational
field is non potential. They proved that the time gap 
for clock synchronization in the external space with respect to 
synchronization through the wormhole handle grows with time. 
The wormhole becomes a time machine 
as soon as this time gap is larger than the time
needed for light propagation between the two mouths in the external space.
In our case the creation of a time machine can be seen explicitly by    
studying the coordinate transformations (\ref{tran}, 9).
Notice that these transformations are chosen in such 
a way that the resulting spacetime is symmetric with respect to the 
spacelike hypersurface $T=t=0$. 
We see that any other slice of constant external time $T$ which enters mouth
$S_1$ will reemerge at mouth $S_2$ at a different time $T'=CT$.
Hence, there exists a time gap $\triangle T_{gap}$ between 
the wormhole mouths which is changing linearly in $T$:  
\begin{equation}\label{gap}
\triangle T_{gap}= \vert T'-T \vert = \vert (C-1)T \vert 
\end{equation} 
For the case that the wormhole mouths are kept in equilibrium by strings
extending to infinity (independent of the choice of the original (10)
or the perturbed metric (27)) the external time interval
$\triangle T_{NG}$ needed by a null geodesic propagating from the inner
pole $A$ of wormhole mouth $S_1$ to the inner pole $B$ of mouth $S_2$ is
given by:
\begin{eqnarray}
\triangle T_{NG}=D-R(e^{\delta(0)}+e^{\epsilon(0)})+b_2+b_1 \nonumber\\
+2b_1\left(\frac{D-b_1+b_2}{D-b_1-b_2}\right)
\log{\left(\frac{D-Re^{\delta(0)}+b_2-b_1}
{Re^{\delta(0)}-2b_1}\right)}\nonumber\\
+2b_2\left(\frac{D-b_2+b_1}{D-b_1-b_2}\right)
\log{\left(\frac{D-Re^{\epsilon(0)}+b_1-b_2}{Re^{\epsilon(0)}-2b_2}\right)}
\end{eqnarray} 
Because this is the shortest path between the two wormhole mouths, 
CTCs exist if the time gap $\triangle T_{gap}$ is larger
than this time intervall $\triangle T_{NG}$. 
By symmetry with respect to the hypersurface $T=t=0$ 
the constructed spacetime consists of two regions 
containing CTCs separated by a region without CTCs.
If for example $b_2 > b_1$ and therefore $C>1$, 
a causality horizon will appear as soon as 
the null geodesic running along
the $z-axis$ from point $A$ to $B$ is closed. This happens at the value  
of exterior time $T_{CH}$, where $T_{CH}$ is given by
\begin{equation}
\triangle T_{NG}=\triangle T_{gap}=: (C-1) T_{CH}.
\end{equation}

\section*{Acknowledgements}
This work was initiated at the workshop on "Mathematical Relativity" held
at the Erwin-Schr\"odinger-Institute for Mathematical Physics (ESI). 
W. Israel thanks the ESI for hospitality. 
We also acknowledge support from the FUNDACION FEDERICO.

\section*{Appendix A}
We prove that for large D there exists a deformation function 
$\epsilon(\vartheta)$ which satisfies equation (\ref{ns}):
\begin{equation} 
F_D(\vartheta,e^{\epsilon}):=e^{\epsilon}-e^{\lambda_{1}(\epsilon,\vartheta)-
\lambda_{2}(\delta,\vartheta)}e^{\delta}=0
\end{equation} 
Notice that $e^{\delta}$ is a function of $e^{\epsilon}$ given by 
eqn. (\ref{delta}) and
that $\lambda_1$ is to be taken on $S_2$ while $\lambda_2$ on $S_1$:
\begin{equation}
e^{\lambda_{1}(\epsilon,\vartheta)-\lambda_{2}(\delta,\vartheta)}
=\sqrt{\left(1-\frac{4b_1}{\rho_1^++ \rho_1^-+2b_1}\right)
       \left(1+\frac{4b_2}{\rho_2^++ \rho_2^--2b_2}\right)}
\end{equation}
\begin{eqnarray}
\rho_1^{\pm} \vert_{S_2}=\{Re^{\epsilon}(Re^{\epsilon}-2b_2)\sin(\vartheta)^2+
                  (D-(Re^{\epsilon}-b_2)\cos(\vartheta) \mp b1)^2\}^{\frac12}
              \nonumber\\
\rho_2^{\pm} \vert_{S_1}=\{Re^{\delta}(Re^{\delta}-2b_1)\sin(\vartheta)^2+
                  (D-(Re^{\delta}-b_1)\cos(\vartheta) \pm b_2)^2\}^{\frac12}
\end{eqnarray}
We show that $e^{\epsilon(\vartheta)}$ can be 
approximated by using Newton's method and additionally that its derivative 
$\frac{d}{d\vartheta}e^{\epsilon(\vartheta)}$ exists and can be 
calculated by implicit differentiation of $F_D(\vartheta,e^{\epsilon})$.
  
Since for $b_1=b_2$ the function 
$F_D(\vartheta,e^{\epsilon})$ vanishes identically,
we assume - without loss of generality - that $b_2>b_1 \geq 0$. 
Under this condition $F_D(\vartheta,e^{\epsilon})$ satisfies
the following properties which tell us  
that for every fixed $\vartheta$ there exists at least one zero on the
interval $e^{\epsilon} \in I=[\frac{2b_2}{R},1]$ :
\begin{equation}\label{prop1}
F_D(\vartheta,\frac{2b_2}{R})=\frac{2b_2}{R}-
e^{\lambda_{1}-\lambda_{2}}\frac{2b_1}{R}>0
\end{equation}
\begin{equation}\label{prop2}
F_D(\vartheta,1)=1-e^{\lambda_{1}-\lambda_{2}}<0
\end{equation}   
These two inequalities can easily be verified for sufficient large $D$ 
by expanding $e^{\lambda_1-\lambda_2}$ in powers of $\frac{R}{D}$,
\begin{equation}
e^{\lambda_1-\lambda_2}(\frac{R}{D})
= 1+\frac{b_2-b_1}{D}+\frac12\frac{d^2}{(d\frac{R}
{D})^2}e^{\lambda_1-\lambda_2}(\xi) 
\left(\frac{R^2}{D^2}\right) \quad \mbox{, where} \quad \xi\in [0,\frac{R}{D}]
\end{equation}
By estimating the remainder of the expansion we actually can show that
properties (\ref{prop1}) and (\ref{prop2}) hold for at least $D>4R$.
	 
For large $D$ the function $F_D(\vartheta,e^{\epsilon})$ 
is strictly monotonically decreasing and concave with respect to $e^{\epsilon}$.
This can be seen by
considering $F_D(\vartheta,e^{\epsilon})$ and its first and second derivative 
with respect to $e^{\epsilon}$ as a sequence of functions labeled by $D$. 
For every fixed $\vartheta$ this sequence converges uniformly on $I$ to a limit 
function which can easily be estimated:
\begin{equation}\label{prop3}
\lim_{D \to \infty} \frac{d}{de^\epsilon} F_D(\vartheta,e^{\epsilon})=
1-C^2\frac{(e^\epsilon-\frac{b_2}{R})}{(e^\delta-\frac{b_1}{R})} \leq
-\frac{b_2-b_1}{R}\frac{1}{(1-\frac{2b_2}{R})(1-\frac{b_1}{R})}
\quad \mbox{on I}
\end{equation}
\begin{equation}\label{prop4}
\lim_{D \to \infty} \frac{d^2}{(de^\epsilon)^2} F_D(\vartheta,e^{\epsilon})
<0  \quad \mbox{on I}
\end{equation}
A general theorem about the convergence of Newton's method 
(see for example Heuser I, p.408) says that if
conditions (\ref{prop1},\ref{prop2},\ref{prop3} and \ref{prop4}) hold 
Newton's sequence, defined by
\begin{equation}
e^\epsilon_{n+1}:=
e^\epsilon_n-\frac{F_D(\vartheta,e^\epsilon_n)}
{\frac{d}{de^{\epsilon}}F_D(\vartheta,e^\epsilon_n)}\quad,
\end{equation}
converges to a unique zero $\zeta$ in $I$ when started from $e^\epsilon_0=1$.
We perform one iteration and estimate the error by  
\begin{equation}
\vert e^\epsilon_{1}-\zeta \vert \leq
\frac{\vert F_D(\vartheta,e^\epsilon_1)\vert}{\mu}
=C_1\left(\frac{R}{D}\right)^2
\quad \mbox{where}\quad
\mu:=\min_{e^\epsilon\in I} \vert \frac{d}{de^{\epsilon}}
F_D(\vartheta,e^\epsilon)\vert
\end{equation}
The existence
of such a constant $C_1$ for large $D$ follows from the 
facts that $\mu$ is bounded 
away from zero (see eqn.(\ref{prop3})) for $D$ tending to infinity 
and that the first two terms of the expansion of the function 
$F_D(\vartheta,e^\epsilon_1)$ in powers of $\frac{R}{D}$ vanish.
Without exactly determining $C_1$
we may conclude that 
$e^\epsilon_{1}$ (eqn.(\ref{eps})) gives
the zero $\zeta$ correctly to order $\frac{R}{D}$.

It is easy to see that the partial
derivative $\frac{\partial}{\partial \vartheta} F_D(\vartheta,e^{\epsilon})$
exists and on the other hand we have shown that for sufficient large $D$ the
partial derivative  
$\frac{d}{de^\epsilon} F_D(\vartheta,e^{\epsilon})$ exists and is nonzero
(see eqn.(\ref{prop3})). Therefore, it follows from the implicit 
function theorem 
that $e^{\epsilon(\vartheta)}$ is differentiable and $\frac{d}{d\vartheta}
e^{\epsilon(\vartheta)}$ is given by
\begin{equation}
\frac{d}{d\vartheta} e^{\epsilon(\vartheta)} =
-\frac{\frac{\partial}{\partial \vartheta} F_D^{\vartheta}(e^{\epsilon})}
{\frac{d}{de^\epsilon} F_D^{\vartheta}(e^{\epsilon})}.
\end{equation} 
Expanding this quantity in powers of $\frac{R}{D}$ leads to the 
estimate given by equation (\ref{deriv}).

\begin{samepage}
\section*{Figure}
\begin{center}
\leavevmode
\epsfxsize \textwidth \epsfbox{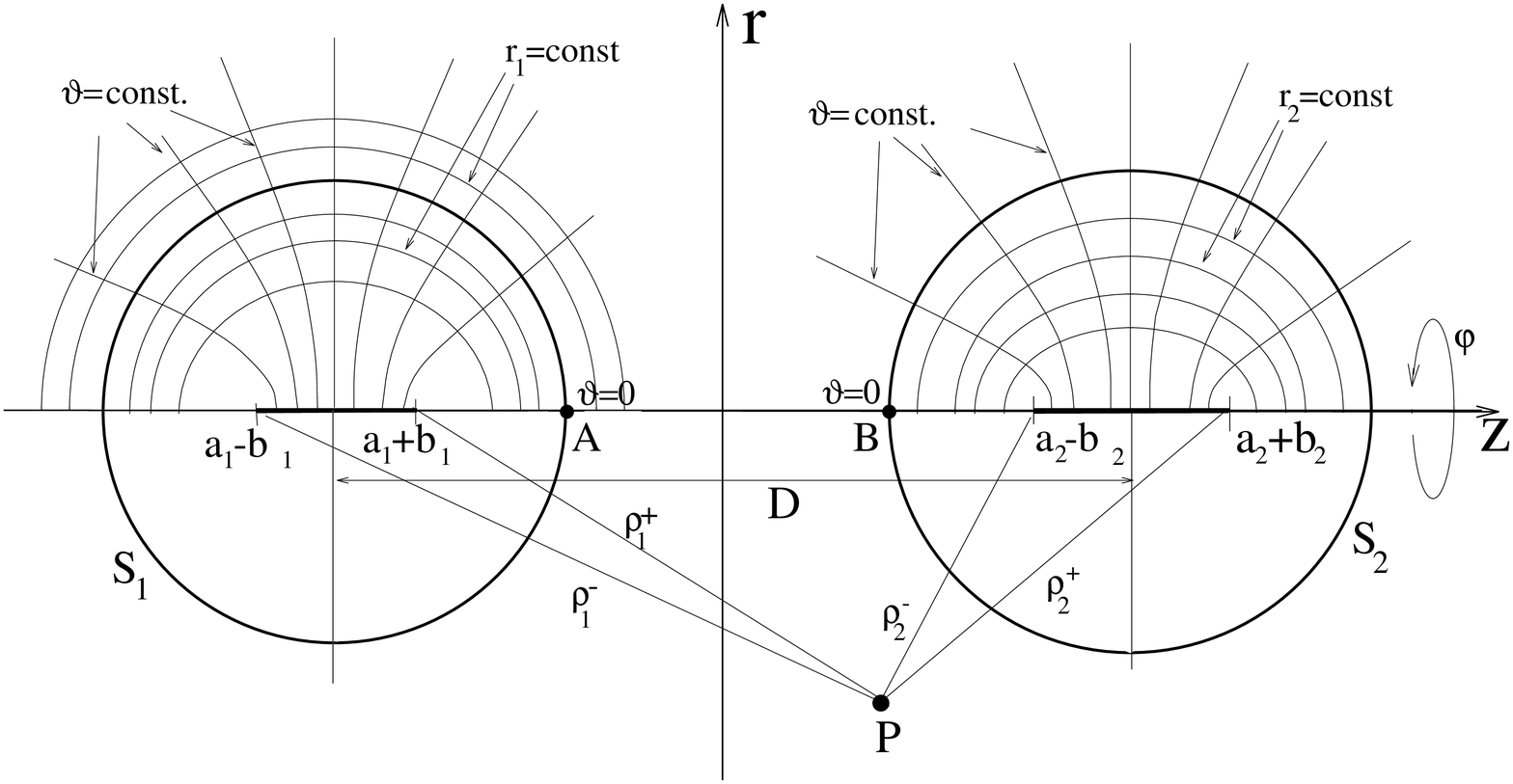}
\end{center}
\renewcommand{\baselinestretch}{1}
\small \normalsize
{\bf Figure 1:} {\small The relation between the cylindrical
coordinate system $\tilde
I(T,r,z,\varphi)$ and the two spherical coordinate systems 
$I(t,r_1,\vartheta,\varphi)$ and
$II(t,r_2,\vartheta,\varphi)$ centered at the wormhole mouths $S_1$ and
$S_2$ respectively is shown. 
The time and angle coordinates $(T,t,\varphi)$ 
are suppressed and we have not distinguished angle coordinates
$(\vartheta, \varphi)$ of the different patches $\tilde I$, $I$ and $II$.
The figure also makes clear the 
geometrical meaning of the quantities $\rho_i^{\pm}$ defined in Section 2.} 
\end{samepage}    	

\end{document}